\documentstyle[12pt]{article}

\newcommand{\be}{\begin{equation}}
\newcommand{\ee}{\end{equation}}

\newtheorem{teo} {Theorem} [section]
\newtheorem{defi}[teo] {Definition} 
\def\bkC{{\rm \kern.24em 
\vrule width.05em 
\vrule height1.4ex 
depth-.05ex 
\kern-.26em C}}

\begin{document}

\setlength{\oddsidemargin}{0cm} \setlength{\baselineskip}{7mm}

\begin{normalsize}\begin{flushright}
\end{flushright}\end{normalsize}

\begin{center}
  
\vspace{25pt}
  
{\Large \bf Simplicial minisuperspace models in the presence of a
massive scalar field with arbitrary scalar coupling $\eta R\phi^{2}$ }

\vspace{40pt}
  
{\sl Crist\'{o}v\~{a}o Correia da Silva}
$^{}$\footnote{e-mail address :
clbc2@damtp.cam.ac.uk}
and {\sl Ruth M. Williams}
$^{}$\footnote{e-mail address : rmw7@damtp.cam.ac.uk}
\\

\vspace{20pt}

DAMTP, Silver Street, Cambridge, \\
CB3 9EW, England \\

\end{center}

\vspace{40pt}

\begin{center} {\bf ABSTRACT } \end{center}
\vspace{12pt}
\noindent

We consider a simplicial minisuperspace model based on a cone over the
$\alpha_{4}$ triangulation of the $3-$sphere, in the presence of a massive
scalar field, $\phi$, with arbitrary scalar coupling term $\eta R \phi^{2}$. By restricting all the interior  edge lengths and all the
 boundary edge lengths to be
 equivalent and the scalar field to be homogenous on the $3-$space, we
 obtain a  two dimensional minisuperspace model $\{s_{i},\phi_{i}\}$ for
what is one of  the most relevant triangulations  of the  spatial universe. 
We solve the two classical equations and find that there are both real
Euclidean and Lorentzian classical solutions for any size of the
boundary $3-$space, $\alpha_{4}$.
After studying the analytic
 properties of the action in the space of complex  edge lengths
 we then obtain   steepest
 descents contours of constant imaginary action passing through
 Lorentzian  classical geometries  giving the dominant contribution and  yielding a convergent wavefunction
 of the universe. We also show that the semiclassical wavefunctions for the
Euclidean solutions associated with large boundary $3-$spaces are
exponentially suppressed,

 Consequently we can be confident that by using the SD contour
 associated with classical Lorentzian solutions  the  semiclassical approximation based on
 those classical solutions is well justified,  clearly  predicting classical spacetime in the
 late universe.  This wavefunction is then evaluated numerically.

\vspace{24pt}

\vfill

\newpage

\section{Introduction}

 The sum over
histories formulation of quantum gravity has undoubtably been one of the
most useful tools in the long running program of quantising
gravity. In this  formulation an amplitude for a
certain state of the universe is constructed by summing over a certain
class of physically distinct histories that satisfy appropriate
boundary  conditions, weighted by their respective  action. There are
no a priori definitions of what the class of histories and boundary
condition should be. As such what we have are proposals for both of
them. In our opinion  Hartle-Hawking's no boundary proposal formulated
on \cite{hawk} is the most natural one. As for the space of
histories, until recently it was widely thought that these histories
should be confined to smooth manifolds with well-behaved
metrics. However, in \cite{witt1} and \cite{witt2}, Schleich and Witt
put forward a  powerful case
for the generalisation of the space of histories to include smooth
conifolds. Computations for concrete models in
\cite{birm}, \cite{witt3} and by us \cite{cris1} have reinforced that proposal.

However, even this generalisation of the Hartle-Hawking proposal is plagued by at least two main
problems. Firstly, the Euclidean gravitational action is not bounded
from below which leads to the divergence of the Euclidean path
integral. Secondly,  there is no clear prescription for the
correct integration contour to use. In  \cite{h2}, Hartle proposed
the use of the steepest descents contour in the space of complex
metrics as the solution to both problems. Furthermore 
by choosing  the steepest descents contour passing through the classical
solutions of the theory, he  made it very likely that the path
integral be dominated by classical four-geometries, i.e., solutions of
Einstein's equations and stationary points of the path integral, as
desired for any wave function that is intended to represent our current Universe. 
Note that in this view, the fact that an integration solely over
real-valued Euclidean geometries does not yield a convergent result
for the path integral, is actually a good thing, for such a path integral would never predict the oscillatory behaviour in the late Universe that  traditionally represents classical Lorentzian space-time.

However, the usefulness of the above formulation in computational
terms leaves something to be desired. The resulting functional
integrations over the metric tensor and matter fields are usually very
hard, no matter what integration contour we choose. It is here that a
simplicial formulation can be of great help. As we shall see below, in
less than seven dimensions
smooth manifolds are in a one-to-one correspondence with a special
kind of simplicial complexes called combinatorial
manifolds. This means that we can
substitute a sum over smooth manifolds and metric tensors by a sum
over simplicial complexes and their squared edge lengths, which in
certain cases greatly simplifies calculations. However, as in the
continuum case, a complete sum is still very hard, so we also end up
using approximate minisuperspace models with features that greatly reduce
the complexity of the calculations, in particular simplicial  minisuperspace models  based in Regge
calculus. Such simplicial minisuperspace models were introduced by
Hartle \cite{h1}.  One typically takes the simplicial
complex which models the topology of interest to be fixed and the
square edge length assignments play the role of the metric degrees of
freedom. The summation over edge lengths models the continuum
integration over the metric tensor. This approach has several
advantages. First by treating the four-geometry directly it is more
adequate to deal with the Hartle-Hawking proposal, \cite{hawk}, (with its four-dimensional nature), than the usual $3+1$ ADM decomposition of space-time, where a careful study of how the four-geometry closes off at the beginning of the universe is essential. 
Second, by discretizing space-time the classical equations become
algebraic which makes it easier to find classical solutions which are
essential to the semiclassical approximation. Third, the simplicial
minisuperpspace models  offer the possibility of systematic improvement.

In the present case we consider one such model where the topology has
been fixed to that of a cone over $\alpha_{4}$, which is  the simplest
triangulation of the $3-$sphere. As a matter sector we consider a
massive scalar field with arbitrary scalar coupling to gravity, $\eta R\phi^{2}$.

Unlike the results of our previous papers \cite{cris1} and \cite{cris2}
where the only real classical solutions for universes with large $3-D$ boundary,(like our own), were Lorentzian solutions, the introduction
of the scalar coupling leads to the additional existence of real
Euclidean solutions for universes with boundary of arbitrarily large  size. However, the
semiclassical wavefunction associated with such solutions proves to be
exponentially suppressed and so the Lorentzian solutions still
dominate, as they should if our model is to predict classical
spacetime for the late universe.

\section{Simplicial Framework}

The crucial point in implementing any sum over histories formulation
of quantum gravity is the specification of the set of histories to be
considered. A history in quantum gravity is specified by its topology,
smooth structure and geometry. In the case of the Hartle-Hawking
approach,  the histories considered have been topological spaces with
the topology and smooth structure of a smooth compact manifold and a
geometry specified on that manifold. Lately several papers like \cite{witt3}
and \cite{cris1},
have pointed to the advantages of extending such space of histories
to include histories having more general topology, namely
conifolds. In this paper we shall not deal with this issue.

The traditional choice of histories described above derives from the
fact that, in the (Euclideanized) classical theory of gravity a
classical history is a Riemannian manifold $(M^{n},A,g)$, where $g$ is a Riemannian metric and:

\begin{defi}

The pair $(M^{n},A)$, where $A=\{(U_{a},\Phi_{a})\}$, is a smooth
manifold with atlas $A$ if it satisfies the
following conditions:

\begin{itemize}

\item Every point of $M^{n}$ has a neighbourhood $U_{a}$ which is
homeomorphic to an open subset of $R^{n}$, via a mapping 
$$\Phi_{a}:U_{a}\rightarrow  R^{n}$$.

\item Given any two neighbourhoods with nonempty intersection, then the
mapping

$$\Phi_{b}\Phi_{a}^{-1}:\Phi_{a}(U_{a}\cap U_{b})\rightarrow \Phi_{b}(U_{a}\cap U_{b})
$$
is a smooth mapping between subsets of $R^{n}$.

\end{itemize}
\end{defi}

A topological space that satisfies only the first condition is called
a topological manifold. Such spaces are not appropriate as histories
because the lack of a smooth structure, i.e., atlas, makes it
impossible to define essential concepts on them, like distance,
continuous and 
differentiable functions (like scalar fields), integration, etc.

The concrete implementation of a sum over smooth manifolds is very
difficult. One of the main problems is how to provide a finite
representation for the manifold-based histories. A simplicial
formulation of quantum gravity aims to provide one such
representation. For that to happen one must prove that there is a one
to one correspondence between the set of smooth manifolds and some set
of simplicial complexes.

A simplicial complex somehow plays the role of topological manifold in
the simplicial framework, in the sense that it also lacks the necessary structure to define
essential concepts like dimension, distance, volume, curvature etc.

\begin{defi}

 A simplicial complex $(K,\mid K\mid )$ is a topological space $\mid K\mid$ and a
 collection of simplices $K$, such that

\begin{itemize}

   \item $\mid K\mid$ is a closed subset of some finite dimensional
   Euclidean space.

   \item  If $\sigma $ is a face of a simplex in $K$, then $\sigma $
   is also contained in $K$.

   \item  If $\sigma _{a}$ and $\sigma _{b}$ are simplices in $K$, then  
   $\sigma _{a} \cap \sigma _{b}$ is a face of both  $\sigma _{a}$
   and $\sigma _{b}$.

   \item The topological space $\mid K\mid $ is the union of all
   simplices in $K$.
\end{itemize}     
\end{defi}

If we are to be able to define essential concepts such as continuity and
differentiability of functions on simplicial complexes we will need
 to introduce some kind of structure similar to that  of
smooth manifolds.
 To do so a few more definitions are
necessary. 

Remember that the PL-join of a point $a$ with a set $L$, denoted $aL$, is
the union of all line segments joining points of $L$ with the point
$a$. This is also called a PL-cone over $L$ with apex $a$. Remember
also that:

\begin{defi}
A map $f: P\rightarrow Q$ between two polyhedra $P$ and $Q$, is said
to be  piecewise linear,
(PL), if each point $p$ in $P$ has a cone neighbourhood $N=aL$ such
that
$$f(\lambda a+\mu x)=\lambda f(a)+\mu f(x)$$
where $x$ is in $L$, $a$ is the apex of the cone $N$ and $\lambda,\mu
\geq 0, \lambda+\mu=1$.
\end{defi}

\begin{defi}

A PL manifold $M^{n}$ is a topological manifold endowed with a PL-atlas
$A={(U_{a},\Phi_{a})}_{a\in \Lambda}$, such that the mapping 
\be
\Phi_{b}\Phi_{a}^{-1}:\Phi_{a}(U_{a}\cap U_{b})\rightarrow \Phi_{b}(U_{a}\cap U_{b})
\ee

is a piece-wise linear (PL) mapping between subsets of ${R}^{n}_{+}$.
\end{defi}

These PL-manifolds are very closely connected with a special kind of
simplicial complexes, the combinatorial manifolds, which are defined
by imposing additional 
restrictions to the very general definition of simplicial
complex. These restrictions make it possible to define concepts like
distance, volume , curvature , etc.

\begin{defi}
 
A combinatorial $n-$manifold ${\cal M}^{n}$, is an $n-$dimensional simplicial complex such  that  
       
\begin{itemize}

     \item It is pure.
     \item It is non-branching.
     \item Any two $n-$simplices can be connected by a sequence of $n-$simplices, each intersecting along some $(n-1)-$simplex.
     \item  The link of every vertex is a combinatorial $(n-1)-$sphere.

\end{itemize}
\end{defi}

Indeed it can be shown that PL manifolds are equivalent to combinatorial
manifolds, \cite{witt2}.  So every combinatorial
manifold admits a PL-atlas ${(U_{a},\Phi_{a})}_{a\in \Lambda}$,   by which concepts like
continuity and differentiability of scalar fields in combinatorial
manifolds can be defined, thus playing  a similar role to
that of the smooth structure in smooth manifolds.

Furthermore given these definitions, following \cite{witt1}  it can be
proven that in less than seven dimensions every $PL$-manifold has a
$\emph{unique}$ smoothing so we can state a very important result,
namely:

\vspace{20pt}

   In less than seven dimensions, every smooth manifold, $M^{n}$, is triangulated by a ${\it{unique}}$ combinatorial manifold, ${\cal M}^{n}$. 
\vspace{20pt}

Obviously  each smooth  manifold has several distinct
triangulations, what this result says is that  only one of them is a
combinatorial triangulation, i.e., a triangulation based on a
combinatorial manifold, and not just any simplicial complex.

So we see that the topological part of the sum over histories 
 can be recast in terms of simplicial representatives of
the ``continuum'' spaces. However in order to define a concrete sum over simplicial histories we still need to associate a metric and an action to each simplicial complex to be considered. Note that up to now we have not specified any kind of metric information associated with simplicial complexes. Once we have fixed the topology of the underlying simplicial complex, the most convenient way to attach metric information to it, is to use Regge calculus.

\subsection{Regge calculus}

A convenient way of defining an $n-$simplex is to specify the
coordinates of its $(n+1)$ vertices, $\sigma =[0,1,2,...,n]$. By
specifying the squared values of the lengths of the edges $[i,j]$,
$s_{ij}$, we fix the simplicial metric on the simplex:

\be
\label{simpmet}
g_{ij}(s_{k})=\frac{s_{0i}+s_{0j}-s_{ij}}{2}
\ee
where $i,j=1,2,..n$.

So if we triangulate a smooth manifold $M$ endowed with a metric $g_{\mu \nu}$ by a homeomorphic simplicial manifold 
${\cal{M}}$, the metric information is transferred to the simplicial metric of that simplicial complex

\be 
g_{\mu \nu}(x) \longrightarrow   g_{ij}(\{s_{k}\})=\frac{s_{0i}+s_{0j}-s_{ij}}{2}
\ee

In the continuum framework the sum over metrics is implemented through a functional integral over the metric components $\{g_{\mu \nu}(x)\}$. In the simplicial framework the metric degrees of freedom are the squared edge lengths, and so the  functional integral is replaced by a simple multiple integral over the values of the edge lengths. But not all edge lengths have equal standing. Only the ones associated with the interior of the simplicial complex get to be integrated over:

\be
\int Dg_{\mu \nu }(x) \longrightarrow \int D\{s_{i}\}=\prod \int d\mu (s_{i}) 
\ee

The boundary edge lengths remain after the sum over metrics and become the arguments of the wavefunction of the universe.
In the simplicial framework the fact that the geometry of the
 complexes is completely fixed by the specification of the squared
 values of all edge lengths, means that all geometrical quantities,
 such as volumes and curvatures, can be expressed completely in terms
 of those edge lengths.

We shall also be considering a scalar field with arbitrary mass and
scalar curvature coupling, taking values $\phi_{k}$ at each vertex $k$ of the
complex. Generally these values will be 
independent, but like the edge lengths, not all have equal
standing. Only the ones associated with interior vertices, $\{\phi_{i}\}$ are to be
integrated over:

\be
\int d\phi \longrightarrow \int D\{\phi_{i}\}=\prod \int d\phi_{i} 
\ee
The values of the field at the boundary vertices
$\{\phi_{b}\}$ are just boundary conditions, becoming the arguments of
the wavefunction.

The Euclideanized Einstein action for a smooth $4-$manifold $M$ with
boundary $\partial M$, and endowed with a $4-$metric, $g_{\mu \nu }$,
and a scalar field $\phi $ with arbitrary mass $m$ and scalar
curvature coupling constant $\eta$ is 

\begin{eqnarray*}
I[M,h_{ij},\phi ]&=&-\int _{M}d^{4}x\sqrt{g}\frac{(R-2\Lambda
                       )}{16\pi G}-\int _{\partial
                       M}d^{3}x\sqrt{h}\frac{K}{8\pi G}+ \\
                       &+& \mbox{} \frac{1}{2}\int
                       _{M}d^{4}x\sqrt{g}(\partial _{\mu}\phi \partial
                       ^{\mu} \phi +m^{2} \phi ^{2}+\eta R\phi^{2})   
\end{eqnarray*}
where $K$ is the extrinsic curvature.

Its simplicial analogue will be the Regge action for a combinatorial
$4-$manifold, ${\cal {M}}$, with squared edge lengths $\{s_{k}\}$, and
with a scalar field taking values $\{\phi _{v}\}$ for each vertex $v$
of ${\cal {M}}$, \cite{ruth}:

\begin{eqnarray*}
 I[{\cal {M}},\{s_{k}\},\{\phi _{v}\}]&=&\frac{-2}{16\pi G}\sum _{\sigma _{2}^{i}} V_{2}(\sigma _{2}^{i})\theta (\sigma _{2}^{i}) +  {\frac{2\Lambda }{16\pi G}} \sum _{\sigma _{4}}V_{4}(\sigma _{4}) \\
                                      &-&{\frac{2}{16\pi G}}\sum _{\sigma _{2}^{b}} V_{2}(\sigma _{2}^{b})\psi (\sigma _{2}^{b})+\frac{1}{2}\sum _{\sigma _{1}=[ij]} \tilde{V}_{4}(\sigma _{1})\frac{(\phi _{i}-\phi _{j})^{2}}{s_{ij}}\\
                                      &+& \mbox{}\frac{1}{2}\sum
                                      _{j}\tilde{V}_{4}(j)m^{2}\phi
                                      _{j}^{2}+  \frac{1}{2}\sum
                                      _{j}\tilde{V}_{4}(j)\eta R_{j}\phi _{j}^{2} 
\end{eqnarray*}
where:

\begin{itemize}

 \item $\sigma _{k}$ denotes a $k-$simplex belonging to the set $\Sigma _{k}$ of all  $k-$simplices in  ${\cal {M}}$.

\item $\theta(\sigma _{2}^{i})$, is the deficit angle associated with the interior $2-$simplex $\sigma _{2}^{i}=[ijk]$

\be
  \theta(\sigma _{2}^{i})=2\pi -\sum _{\sigma _{4}\in St(\sigma _{2}^{i})}\theta _{d}(\sigma _{2}^{i},\sigma _{4})   
\ee

and $\theta _{d}(\sigma _{2}^{i},\sigma _{4})$ is the dihedral angle between the $3-$simplices $\sigma _{3}=[ijkl]$ and $\sigma ^{'}_{3}=[ijkm]$, of $\sigma _{4}=[ijklm]$ that intersect at  $\sigma _{2}^{i}$. Its full expression is given in \cite{birm}.

\item   $\psi (\sigma _{2}^{b})$ is the deficit angle associated with the boundary $2-$simplex $\sigma _{2}^{b}$:

\be
  \psi (\sigma _{2}^{b})=\pi -\sum _{\sigma _{4}\in St(\sigma _{2}^{b})}\theta _{d}(\sigma _{2}^{b},\sigma _{4})    
\ee

\item $V_{k}(\sigma _{k})$ for $k=2,3,4$ is the $k-$volume associated with the $k-$simplex, $\sigma _{k}$, and once again their explicit expressions in terms of the squared edge lengths are given in  \cite{birm}.

\item  $\tilde{V}_{4}(\sigma _{1})$, is the $4-$volume in the simplicial complex ${\cal{M}}$, associated with the edge $\sigma _{1}$, i.e., the volume of the space occupied by all points of ${\cal{M}}$ that are closer to $\sigma _{1}$ than to any other edge of  ${\cal{M}}$. The same holds for $\tilde V_{4}(j)$ where $j$ represents all vertices of ${\cal{M}}$.

\item  It can be shown, \cite{ruth}, that

$$\tilde{V}_{4}(j)R_{j}=\frac{2}{3}\sum _{\sigma_{2}[jkl]}V_{2}(\sigma_{2})\chi(\sigma_{2})$$

where the sum is over all triangles $\sigma_{2}$ that contain the
vertex $j$, and $\chi(\sigma_{2})$ is the deficit angle associated
with  $\sigma_{2}$. We use the new symbol $\chi$ because the triangle
$\sigma_{2}$ can be an interior  or boundary triangle.

\end{itemize}

It is easy to see that both $\tilde{V}_{4}(\sigma _{1})$ and $\tilde
V_{4}(j)$, can be expressed exclusively in terms of the edge lengths
$\{s_{k}\}$. In fact all the previous terms can be written exclusively
in terms of $\{s_{k}\}$ and $\{\phi_{k}\}$.

So we see that any  history in QG of the type
$(M^{4},A,g_{\mu \nu},\phi )$, where $M^{4}$ represents  a topological
manifold  endowed with a smooth structure $A$, metric $g_{\mu \nu}$,
and in the presence of matter fields represented by $\phi$, has an
{\it{unique}} simplicial analogue, $({\cal{M}}^{4},\{s_{k}\},\{\phi
_{j}\})$. This allows us to concretely implement the formal sum over
histories in terms of this finite representation as:

\be
\label{2}
\Psi [\partial {\cal{M}},\{s_{b}\},\{\phi _{b}\}]=\sum _{{\cal{M}}^{4}}\int D\{s_{i}\}D\{\phi _{i}\}e^{-I[{\cal{M}}^{4},\{s_{i}\},\{s_{b}\},\{\phi _{i}\},\{\phi _{b}\}]}  
\ee
where

\begin{itemize}

\item $ \{s_{i}\}$ are the squared lengths of the interior edges 

\item  $\{s_{b}\}$ are the squared lengths of the boundary  edges 

\item  $\{\phi_{i}\}$ are the values of the field at the interior vertices 

\item  $\{\phi_{b}\}$ are the values of the field at the boundary vertices

\end{itemize}

Although the functional integral over metrics has been written explicitly
 in terms of the edge lengths, this expression is still
heuristic because we still need to specify  the list of suitable
simplicial complexes ${\cal{M}}^{4}$ we intend to sum over,  the
measure,  and the integration contour to be used. To circumvent these
 problems we shall compute the sum approximately by singling out a 
subfamily of simplicial histories described by only a few parameters
and carrying out the sum over these histories alone.

 An example of this is to adopt a simplicial minisuperspace approximation. We now describe in some detail the minisuperspace model we shall consider.

\section{Simplicial Minisuperspace}

We shall reduce our attention to a significant  subfamily of simplicial histories characterised by the following restrictions:

 We shall consider that the universe has only one $S^{3}$
 boundary and it is well approximated as a
 simplicial cone over the closed combinatorial $3$-manifold
 $\alpha_{4}$, which is the 
  simplest  triangulation of the $3-$sphere, $S^{3}$. 

\be   
{\cal {M}}^{4}=a*{\alpha_{4}}
\ee

The combinatorial manifold $\alpha_{4}$ has been described in detail
elsewhere \cite{h1}. We can see a representation of it in
figure $1$. It has $5$ vertices, each connected to all others. 

Note that since all vertices of ${\cal {M}}^{4}$, (even the interior
one) have combinatorial links that are homeomorphic to a $3-$sphere,  ${\cal {M}}^{4}$
is a combinatorial $4-$manifold.

\begin{itemize}

\item  By using a cone-like structure, translated by
the existence of only one interior vertex, the apex $a$ which we shall
henceforth denote as $0$, the only boundary of ${\cal {M}}^{4}$
is $\alpha_{4}$ . So it is very easy to define boundary simplices
and interior simplices. If a simplex contains the interior vertex $0$
it is an interior simplex if not it is a boundary simplex. Moreover,
all interior $p-$simplices in  ${\cal {M}}^{4}$ are cones over
$(p-1)-$simplices of  $\alpha_{4}$ with apex $0$.

\item If we label the five  boundary vertices of ${\cal {M}}^{4}$ simply as $1,2,3,4,5$, then the cone-like structure of ${\cal {M}}^{4}$ leads to all interior edges being of the same form $[0,b]$, with $b =1,2,3,4,5$.
So it makes sense to introduce the restriction that all interior edges
have equal lengths whose squared value is denoted $s_{i}=s_{0b }$. A
similar assumption is made with respect to the boundary edge lengths,
i.e., we consider them all to be equal to a common value
$s_{ij}=s_{b}$, with $i,j=1,2,3,4,5$. We thus obtain an
$\emph{isotropic and homogeneous }$ triangulation of the $4-$universe.
This leads to an
enormous simplification in the metric part of the integration for the
wavefunction $(\ref{2} )$, since the multiple integral $\int D\{s_{i}\}$ is reduced to a single integral $ \int ds_{i}$. It also greatly simplifies the expression of the simplicial action since there will only be one type of boundary and interior triangle.

\item The simplifications assumed with respect to the edge lengths make it
natural to assume that the scalar field is spatially homogeneous and
isotropic. So we assume that the scalar field takes the same value
$\phi _{b}$ for all boundary vertices of ${\cal {M}}^{4}$. The value
at the interior vertex, $\phi _{i}$, is independent. Again, this leads to an
 simplification in the matter fields  part of the integration for the
wavefunction $(\ref{2})$, since the multiple integral $\int
D\{\phi_{i}\}$is reduced to a single integral $ \int d\phi_{i}$.

\end{itemize}

\subsection{Minisuperspace Wavefunction}

We can now concretely implement a simplicial minisuperspace approximation to the wavefunction of the universe of the type $(\ref{2})$, as

\be
\label{ppsi}
\Psi [\alpha_{4},s_{b},\phi _{b}]=\int ds_{i}d\phi _{i}e^{-I[\alpha_{4},s_{i},s_{b},\phi _{i},\phi _{b}]}   
\ee

The Regge action for this minisuperspace can now be calculated. For simplicity we introduce rescaled metric variables:

\be
 \xi =\frac{s_{i}}{s_{b}}
\ee
\be
 S=\frac{H^{2}s_{b}}{l^{2}}
\ee
where $H^{2}=l^{2}\Lambda /3$, and $l^{2}=16\pi G$ is the Planck length. We shall work in units where $c=\hbar =1$.

Then the volume of the $4-$simplices in  ${\cal {C}}^{4}$ is

\be
 V_{4}(\sigma _{4})=\frac{l^{4}}{24\sqrt{2}H^{4}}S^{2}\sqrt{\xi -3/8}.
\ee

The volume of the $10$ internal $2-$simplices, $\sigma _{2}^{i}$ in ${\cal {C}}^{4}$ is

\be
V_{2}(\sigma _{2}^{i})=\frac{l^{2}}{2H^{2}}S\sqrt{\xi -1/4}.  
\ee

The volume of the $10$ boundary $2-$simplices, $\sigma _{2}^{b}$ in  ${\cal {C}}^{4}$ is

\be
V_{2}(\sigma _{2}^{b})=\frac{\sqrt{3}l^{2}}{4H^{2}}S. 
\ee

The volumes of the internal and boundary $3-$simplices of ${\cal {M}}^{4}$ are, respectively
\be
  V_{3}(\sigma _{3}^{i})=\frac{l^{3}}{12H^{3}}S^{3/2}\sqrt{3\xi -1}, 
\ee
\be
  V_{3}(\sigma _{3}^{b})=\frac{\sqrt{2}l^{3}}{12H^{3}}S^{3/2}.   
\ee

There is only one kind of interior $2-$simplex and boundary  $2-$simplex
The dihedral angle associated with each interior $2-$simplex is

\be
\theta (\sigma _{2}^{i})=\arccos{\frac{2\xi -1}{6\xi -2}}.  
\ee

for the boundary $2-$simplices we have  

\be
\theta (\sigma _{2}^{b})=\arccos {\frac{1}{2\sqrt{6\xi -2}}}.  
\ee

With respect to the matter terms, the kinetic term vanishes when the
edges $\sigma _{2}$ are boundary edges. The only non-vanishing
contribution comes from the internal edges $\sigma _{2}=[0j]$.

Computing the relevant volumes associated with the internal edges and
all the vertices we conclude that the  Regge action for
this simplicial minisuperspace is

\begin{eqnarray*}
I[\xi ,S,\phi _{i},\phi _{b}]&=&-\frac{S}{H^{2}}\biggl \{ \biggl(5\sqrt{3}-\frac{5}{2}\sqrt{3}\eta\phi_{b}^{2}l^{2}\biggr)\biggl[\pi -2\arccos {\frac{1}{2\sqrt{6\xi -2}}}\biggr] \\
                                   &+&\biggl[10-\frac{5}{3}\eta \biggl(\phi_{i}^{2}l^{2}+2\phi_{b}^{2}l^{2}\biggr)\biggr] \sqrt{\xi -1/4}\biggl[2\pi -3\arccos{\frac{2\xi -1}{6\xi -2}}\biggr]  \\
                                   &-& \biggl(\frac{1}{24\sqrt{2}}\biggr)\frac{\sqrt{\xi -3/8}}{\xi}(\phi _{i}l-\phi _{b}l)^{2}\biggl\} +\frac{S^{2}}{H^{2}}\biggl\{\biggl(\frac{5}{4\sqrt{2}}\biggr)\sqrt{\xi -3/8}  \\  
                                   &+& \mbox{} \frac{1}{48\sqrt{2}}\biggl(\frac{m^{2}l^{2}}{H^{2}}\biggr)\sqrt{\xi -3/8}(\phi _{i}^{2}l^{2}+4\phi _{b}^{2}l^{2})\biggr\}
\end{eqnarray*}

Note that we will be expressing  the values of the field $\phi $ and its
mass $m$, in Planck units ($l^{-1}$).

Thus  in order to approximate the formal sum over histories by a fully computable expression 

\be
\Psi [\alpha_{4};s_{b};\phi_{b}]=\int _{C}D\xi D\phi _{i}e^{-I [S,\xi ;\phi _{b},\phi _{i}]},  
\ee
we only need to specify the integration contour $C$, and the measure of integration $D\xi D\phi _{i}$.

 As in the previous cases studied by us \cite{cris1}, \cite{cris2},
 in this
 simplified model  the result yielded by a certain contour C is not
 very sensitive to the choice of measure if we stick to the usual
 measures, i.e., polynomials of the squared edge lengths. In our case we take 

\be
 D\xi D\phi _{i}=\frac{ds_{i}}{2\pi il^{2}}d\phi _{i}=\frac{S}{2\pi iH^{2}}d\xi d\phi _{i}  
\ee

Since  in the case of closed cosmologies there is as yet no known
explicit prescription for the integration contour,  one usually takes a pragmatic view, in which we look for contours that lead to the desired features of the wavefunction of the universe. Following \cite{hh}, these features are:

\begin{itemize}

  \item   It should yield a convergent  path integral 

  \item   The resulting wavefunction should predict  classical
  spacetime in the late universe, i.e, oscillating behaviour when the $\Psi $ is well approximated by the semiclassical approximation.

   \item  The resulting wavefunction should obey the diffeo\-mor\-phism con\-straints, in par\-ticular the Wheeler-DeWitt equation.

\end{itemize}

It is well known that any integration contour over real metrics would
yield a wavefunction that does not satisfy any of these basic
requirements. On the other hand, an integration contour over complex
metrics can, if wisely chosen, lead to a wavefunction that does
satisfy them. 

In the simplicial framework, complex metrics arise from complex-valued
squared edge lengths, $(\ref{simpmet})$. The boundary squared edge
length, $S$, has to be real and positive for obvious physical reasons. But the interior squared edge length, $\xi $, can be allowed to take complex values.

Given the above requirements, $\cite{h2}$ proposes that we use a steepest
descents integration contour on the space of complex valued interior
edge-lengths  passing through the classical solutions
 that should dominate the wavefunction in the late universe,
Before we test his proposal it is essential that we do the analytic
study of the action as a multivalued function of the complex variable $\xi$.

\subsection{Analytic Study of the Action}

The action  is trivially analytic with respect to
the variables $\phi _{i},\;\phi _{b}$ and $S$. But its dependence on
the complex-valued $\xi $ is much more complicated. So we shall investigate the analytic properties of $I$ as if it was a function of $\xi $ only, $I=I[\xi ]$, the other variables acting as parameters.

The function $I[\xi ]$ has singularities at $\xi=0$ and $\xi=1/3$, and
square root branch points at $\xi
=1/4,\:1/3$ and $3/8$. These branch points correspond respectively to the vanishing of the volume of the internal $2-$simplices,  $3-$simplices and $4-$simplices. Using 
$$\arccos{z}=-i\log (z+\sqrt{z^{2}-1})$$

we see that $\xi =1/3$ is also a logarithmic branch point, near which
the action behaves like :

\be
I\:\sim i2\biggl(5\sqrt{3}-\frac{5}{2}\sqrt{3}\eta\phi_{b}^{2}l^{2}\biggr)\biggl(\frac{S}{H^{2}}\biggr)\log{(3\xi -1)}
\ee

The multivaluedness of $I[\xi ]$ associated with these branch points
forces us to implement branch cuts in order to obtain a continuous
function. In general, for terms of the type $\sqrt{z-z_{0}}$ we consider a branch cut $(-\infty ,z_{0}]$. So the branch cuts associated with the terms $\sqrt{\xi -3/8}$, $\sqrt{\xi -1/4}$ and $\sqrt{\xi -1/3}$, altogether lead to a branch cut  $(-\infty ,3/8]$. On the other hand, terms of the type $\arccos (z)$ have branch points at $z=-1,+1,\infty $, and usually the associated branch cuts are chosen as $(-\infty ,-1]\cup [1,+\infty )$. These terms are also infinitely multivalued.

The corresponding cuts for the term $\arccos{\frac{2\xi-1}{6\xi -2}}$
are $(\frac{1}{3},\frac{3}{8}]\cup \-[\frac{1}{4},\frac{1}{3})$. On
the other hand, associated with the term $\arccos{\frac{1}{2\sqrt{6\xi -2}}}$ we have one cut  $(\frac{1}{3},\frac{3}{8}]$ associated with $\arccos{u(z)}$, and another $(-\infty, \frac{1}{3}]$ associated with $u(z)=\sqrt{6\xi -2}$.

So when we consider all these branch cuts simultaneously, we see that one way to obtain a continuous action $I$ as a function of $\xi $, is to consider a total branch cut $(-\infty ,\frac{3}{8}]$. Note that this also takes care of the singularity at $\xi =0$.
Although the action  then becomes a continuous function of $\xi $ in
the complex plane with a cut $(-\infty ,\frac{3}{8}]$, it is still
infinitely multivalued. As usual in similar cases, in order to remove
this multivaluedness we redefine the domain where the action is
defined, from the complex plane to the Riemann surface associated with
$I$. The infinite multivaluedness of the action reflects itself in $I$
having an infinite number of branches with different values. The Riemann surface is composed of an infinite number of identical sheets, $\bkC -(-\infty ,\frac{3}{8}]$, one sheet for each branch of $I$.  

We define the first sheet ${\bkC} _{1}$ of $I[\xi ]$ as the sheet where the terms in $\arccos (z)$ assume their principal  values. So the action in the first sheet will be formally equal to the original expression. Note that with the first sheet defined in this way, for real $\xi >3/8$ the volumes and deficit angles are all real, leading to a real Euclidean  action for $\xi \in [\frac{3}{8},+\infty )$ on the first sheet. 
On the other hand, when $\xi $ is real and less than $1/4$ in the first sheet , the volumes become pure imaginary and the Euclidean action becomes pure imaginary. For all other points of this first sheet the action is fully complex.

When the action is continued in $\xi $ once around all finite branch points ($\xi = 1/4,1/3,3/8$), we reach what shall be called the second sheet . It is easy to conclude that the action in this second sheet is just the negative of the action in the first sheet.

Since by $(\ref{simpmet} )$ we see that the simplicial metric in each
$4-$simplex is real $iff$ $\xi $ is real, then the simplicial
geometries built out of these $4-$simplices will be real when  $\xi $
is real. Furthermore the corresponding eigenvalues of $g_{ij}$ are
$\lambda =\{4(\xi -3/8),1/2,1/2,1/2\}$, \cite{birm}. So we see that for real $\xi >3/8$ we have real Euclidean signature geometries, with real Euclidean action, and for real $\xi <1/4$, we have real Lorentzian signature geometries with pure imaginary Euclidean action.

\subsection{Asymptotic Behaviour of the Action}

If we are to compute the integration of $e^{-I}$ along an SD contour,
one of the essential things we have to know is the behaviour of the
integrand, i.e., of the action, at infinity with respect to the
variable $\xi$.  Only then can we be
confident that the integral converges, with the classical solutions
dominating the wavefunction for the late universe.

It is easy to see that as  $ \xi \rightarrow \infty $ the action behaves like

\be
I[\xi ,S,\phi _{i},\phi _{b}]\sim \frac
{\frac{5}{4\sqrt{2}}+\frac{1}{48\sqrt{2}}\biggl(\frac{m}{H}\biggr)^{2}\biggl(\phi_{i}^{2}+4\phi_{b}^{2}\biggr)}{H^{2}} S(S-S_{crit})\sqrt{\xi } 
\ee
where

\be
S_{crit}=\frac{ \biggl[10-\frac{5}{3}\eta
\biggl(\phi_{i}^{2}+2\phi_{b}^{2}\biggr)\biggr] [2\pi -3\arccos
{(1/3)}]}    {\frac{5}{4\sqrt{2}}+\frac{1}{48\sqrt{2}}\biggl(\frac{m}{H}\biggr)^{2}\biggl(\phi_{i}^{2}+4\phi_{b}^{2}\biggr) }  
\label{scrit}
\ee

The asymptotic behaviour of $I$ for large $\xi $ depends on
whether or not $S$ is larger than the critical value
$S_{crit}$. However, contrary to the corresponding model in \cite{cris1} where there was no scalar curvature coupling and the critical value of
$S$ was

\be
S_{crit}^{\eta=0}=\frac{ 10[2\pi -3\arccos
{(1/3)}]}    {\frac{5}{4\sqrt{2}}+\frac{1}{48\sqrt{2}}\biggl(\frac{m}{H}\biggr)^{2}\biggl(\phi_{i}^{2}+4\phi_{b}^{2}\biggr) }  
\ee
and as such could only take values in a limited range, now $S_{crit}$ can  be arbitrarily negative or positive because of its
$\eta$ dependence. This means that the convergence of the integral along
the SD contour, for a given $S$, will  depend on the value of the coupling
constant $\eta$.

\section{Classical Solutions}

The classical simplicial geometries are the extrema of the Regge action we have obtained above. In our minisuperspace model there are two degrees of freedom $\xi ,\phi _{i}$, so the Regge equations of motion will be:

\be
\frac{\partial I}{\partial \xi}=0
\label{classic1}
\ee
and

\be
\frac{\partial I}{\partial \phi _{i}}=0.
\label{cl2}
\ee

They are to be solved for the values of $\xi ,\phi _{i}$, subject to
the fixed boundary data $S, \phi _{b}$. The classical solutions will
thus be of the form $\overline{\xi}(S, \phi _{b})$, and
$\overline{\phi} _{i}(S,\phi _{b})$. The solution $\overline{\xi}(S,
\phi _{b})$ completely determines the simplicial geometry.

Note that we shall be working  on the first sheet. Of course, since on
the second sheet the action is just the negative of this,  the
equations of motion are the same. And obviously every classical
solution   $\overline{\xi}_{I}(S, \phi _{b})$ located on the
first sheet will have a counterpart $\overline{\xi}_{II}$ of the same
numerical  value, but located on the second sheet, and so with an
action of opposite sign, $I[\overline{\xi}_{I}(S, \phi _{b})]=-I[\overline{\xi}_{II}(S ,\phi _{b})]$. So the classical solutions occur in pairs.

The classical equation $(\ref{cl2})$ is

\be
\label{cleqphi}
\phi _{i}=\frac{\phi _{b}}{A(\xi)+\frac{1}{2}\biggl(\frac{m^{2}}{H^{2}}\biggr)\xi S}  
\ee

where
\be
A(\xi)=1+60\eta \xi
\sqrt{2}\sqrt{\frac{\xi-1/4}{\xi-3/8}}\biggl[2\pi -3\arccos{\frac{2\xi-1}{6\xi-2}}\biggr]
\ee

Introducing this equation into the first one $(\ref{classic1})$,  we obtain a very long cubic equation in S for each value
of $\xi$,
given fixed $\eta$ and $\phi_{b}$.

\be
A_{3}(\xi)S^{3}+A_{2}(\xi)S^{2}+A_{1}(\xi)S+A_{0}(\xi)=0,
\ee

where 
\be
A_{3}(\xi)=30\biggl(K^{2}+\frac{2}{15}K^{3}\phi_{b}^{2}\biggr)\xi ^{2},
\ee

\begin{eqnarray*}
A_{2}(\xi)&=&-240\sqrt{2}\biggl(1-\frac{\eta
\phi_{b}^{2}}{3}\biggr)\sqrt{\frac{\xi-3/8}{\xi-1/4}}\biggr[2\pi-3\arccos{\frac{2\xi-1}{6\xi-2}}\biggr]K^{2}\xi^{2}+   \\
          &+& \mbox{} 60\biggl(K+\frac{2}{15}K^{2}\phi_{b}^{2}\biggr)A(\xi)\xi-20\eta
K^{2}\phi_{b}^{2}\frac{\xi^{2}}{\xi-1/3}-\biggl(\xi-\frac{3}{4}\biggr)K^{2}\phi_{b}^{2}
\end{eqnarray*}

\begin{eqnarray*}
A_{1}(\xi)&=&30A(\xi)^{2}\biggl(1+\frac{1}{6}K\phi_{b}^{2}\biggr)-40\eta KA(\xi)\phi_{b}^{2}\frac{\xi}{\xi-1/3}-2K\phi_{b}^{2}[A(\xi)-1]\frac{(\xi-3/4)}{\xi} \\           &-& \mbox{} 480\sqrt{2}\biggl(1-\frac{\eta
\phi_{b}^{2}}{3}\biggr)\sqrt{\frac{\xi-3/8}{\xi-1/4}}KA(\xi)\xi\biggl[2\pi-3\arccos{\frac{2\xi-1}{6\xi-2}}\biggr]+[1-A(\xi)^{2}]K\phi_{b}^{2}
\end{eqnarray*}

\begin{eqnarray*}
A_{0}(\xi)&=&-240\sqrt{2}\sqrt{\frac{\xi-3/8}{\xi-1/4}}\biggl(A(\xi)^{2}-\frac{\eta
\phi_{b}^{2}A(\xi)^{2}}{3}-\frac{\eta
\phi_{b}^{2}}{6}\biggr)\biggl[2\pi-3\arccos{\frac{2\xi-1}{6\xi-2}}\biggr]  \\
          &-& \mbox{} \frac{20\eta \phi_{b}^{2}[A(\xi)^{2}-1]}{\xi-1/3}-[A(\xi)-1]^{2}\phi_{b}^{2}\frac{(\xi-3/4)}{\xi^{2}}
\end{eqnarray*}
with $K=1/2(m/H)^{2}$.

 This equation can then be solved numerically
for $S$, and by inverting the resulting solutions we obtain
several branches of solutions $\xi=\xi _{cl}(S,\phi_{b})$. For
obvious physical reasons we shall accept only solutions with real
positive $S$. In figure $2$ we show such solutions for $\eta =0.015, m=1$,
and  $\phi_{b}=1$. Between the critical points $\xi=1/4$ and  $\xi=3/8$
there are no real solutions. We chose a small value of $\eta $ and included negative
branches so that the
behaviour of the solutions near the critical points is clear. The
behaviour  of the corresponding solutions for larger values of $\eta$ is
of the same type, but the separation between the several branches is
not so obvious. 

It is
easy to see that for any positive value of $S$ there is at least one  classical
Lorentzian solution $(\overline{\xi} < 1/4)$. Furthermore, when $S$ is large (late
Universe) there is only one Lorentzian solution,  $\overline{\xi}_{I}^{L}(S, \phi _{b})$,
(in the first sheet, of course), and is  located near
the critical point $\xi=1/4$, just like the results obtained in
\cite{cris1}. There is also, of course , its counterpart in the second
sheet $\overline{\xi}_{II}^{L}(S, \phi _{b})$, which though numerically equal has symmetrical action. However, the asymptotic behaviour of the classical
solutions is very different from that in \cite{cris1}. Unlike in \cite{cris1}, the Euclidean branch
$(\xi >3/8)$, is not limited to  a finite range of $S$. In the present case the
Euclidean branch goes all the way to $+\infty $, although this is not
at all evident from figure $2$ because it is a large-scale
behaviour, and this is a small-scale picture. We can  see this large-scale behaviour in figure $3$, but
at the cost of the Lorentzian peak at $\xi=1/4$ becoming indistinct
from the imaginary $\xi$ axis. In figure $3$ we
see that for each positive value of $S$ there is always one pair of
Euclidean signature solutions $\overline{\xi}_{I}^{E}(S, \phi
_{b})=\overline{\xi}_{II}^{E}(S, \phi _{b})\in [3/8,+\infty )$, and
$\overline{\xi}^{E}\rightarrow +\infty$ as $S\rightarrow +\infty$

If we
look in the opposite direction, i.e. $\xi \rightarrow -\infty $ another surprise awaits
us. The positive Lorentzian branch eventually becomes negative and
connects with the 
upper negative branch.

So the main difference introduced by the scalar coupling is the
existence of Euclidean $\overline{\xi}_{I}^{E}(S, \phi
_{b})=\overline{\xi}_{II}^{E}(S, \phi _{b})\in [3/8,+\infty )$ solutions for all positive values
of the boundary edge length $S$. In a semiclassical analysis this
seems to mean that the Lorentzian universe can nucleate with arbitrary
size from an Euclidean regime. However the Regge action
grows very fast as  $\xi$ goes from $3/8$ to $+\infty$, as we can see
in figure $4$. Consequently, the Euclidean classical solutions for large
$S$ are very strongly suppressed.

For negative values of $\eta $ the solutions obtained have a very
different behaviour, but they agree with the positive $\eta$ solutions
in some very important points. In order to present these solutions
more clearly we have separated the Lorentzian range $(\xi<3/8)$,
presented in figure $5$, from the Euclidean range $(\xi>3/8)$, in
figure $6$.  As can be seen in figure $5$, where $\eta =-0.15$, despite the
somewhat bizarre behaviour, we still have that for large values of $S$
there is an  unique pair of classical Lorentzian solutions $\overline{\xi}_{I}^{L}(S, \phi
_{b})=\overline{\xi}_{II}^{L}(S, \phi _{b})$ and these will be located near
the critical point $\xi=1/4$. Also, in the Euclidean regime,
$]3/8,+\infty )$, there is an unique pair of Euclidean solutions $\overline{\xi}_{I}^{E}(S, \phi
_{b})=\overline{\xi}_{II}^{E}(S, \phi _{b})\in [3/8,+\infty )$ for any positive value of 
$S$, and those solutions go to $+\infty$ as $S\rightarrow +\infty$, as
shown in figure $6$.

\section{Steepest Descents Contour}

After studying the analytical and asymptotic properties of the action
we can now focus on the Euclidean path integral that yields the wave
function of the Universe $(20)$, $(21)$.

As we have mentioned above there is as yet no universally accepted
prescription for the integration contour $C$ to use in quantum
cosmology. Following Hartle \cite{h2}, we shall accept that the main
criteria any contour should satisfy are that it should lead to a
convergent path integral and to a wave function predicting classical
Lorentzian spacetime in the late Universe. The steepest descents
contour over complex metrics seems to be the leading candidate.   
In the simplicial framework, complex metrics arise from complex-valued
squared edge lengths, $(\ref{simpmet})$.  

We shall look for the steepest descent (SD) contour,  thinking of the
action as a function of only the complex variable $\xi$, and for the
moment consider $\phi_{i}$ to be only a real parameter in $I=I[\xi]$ to be
integrated over later.
In general, an SD contour associated with an extremum ends up either at
$\infty $, at a singular point of the integrand, or at another
extremum with the same value of $Im(I)$. We have seen that  when $S$ is big enough the only classical
solutions are a pair of real Lorentzian solutions $(\overline{\xi
}_{I}^{L}(S,\phi _{b})=\overline{\xi }_{II}^{L}(S,\phi_{b})\rightarrow
1/4)$, and a pair of Euclidean solutions $(\overline{\xi
}_{I}^{L}(S,\phi _{b})=\overline{\xi }_{II}^{L}(S,\phi_{b})>>3/8)$.

In both cases for  each pair of solutions one of these solutions is located on the first sheet and the other on
the second sheet. Thus, they have
pure imaginary actions of opposite sign in the Lorentzian case and
symmetric real actions in the Euclidean case. Given that their actions are
different valued, in  each pair there is no single SD path can go directly from one solution the other
extremum, but it is possible for the two sections to meet in
infinity, and together they form the total SD contour. In effect given that

$$ I[\overline{\xi}]=[I[\overline{\xi}^{*}]]^{*} $$

and $$ I[\overline{\xi}_{I}]=- I[\overline{\xi}_{II}]$$
where $*$ denotes complex conjugation, we see that the SD path that passes through  $\overline{\xi }_{II}$ will be
the complex conjugate of the SD path that passes through
$\overline{\xi }_{I}$
So the \emph{total} SD contour will always be composed of two complex
conjugate sections, each passing through one extremum, and this
together with the real analyticity of the action guarantees that the
resulting wavefunction is real.

 For large $S$ we choose to consider the SD contour
associated with  the Lorentzian solutions for two reasons. First, it
is the only one likely to describe a late universe like our
own. Second, the Euclidean action of the Euclidean solutions
 becomes very large very fast when
$S$ increases, which strongly suppresses these solutions in any
computation of the wavefunction, except when $S$ is small, (see figure $4$).

The  SD path in the complex $\xi$ Riemann surface of $I$, passing
through a generic  classical solution
$\{\xi_{cl}(S,\phi_{b}),\phi_{i}^{cl}(S,\phi_{b})\}$ is defined as:

\be
C_{SD}(S,\phi _{b},\phi_{i})=\biggl\{(\xi \in R) : Im[I(S,\xi ,\phi
_{i},\phi _{b})]=\tilde{I}[\xi_{cl}(S,\phi_{b}),\phi_{i}^{cl}(S,\phi_{b})]  \biggr\}
\label{csd}
\ee
where $R$ is the Riemann sheet of the action, and $\tilde{I}(\xi)=iI(\xi)$.

In figure $7 $ we show the result of a numerical computation of this
path for $m=1$, $\phi_{b}=1$, $\eta=0.015$, $S=150$, and
$\phi_{i}=0.15$. 

The SD path associated with the the other solution in the second sheet
is just the mirror image of this, relative to the real $\xi$ axis.
Together they form the SD contour we are looking for.

 Changes in  the values of
$\phi_{b},\phi_{i}$ and $\eta$, do not alter the generic shape of
this SD contour.It starts at $+\infty$ in the first quadrant and ends in the real $\xi$ axis precisely at the
classical solution to which it is associated.

 Going upward from the Lorentzian  extremum on the first sheet, the SD contour proceeds to
infinity in the first quadrant approximately along the parabola

\begin{eqnarray*}
 & & \biggr[\frac{5}{4\sqrt{2}}+\frac{1}{48\sqrt{2}}\biggl(\frac{m}{H}\biggr)^{2}\biggl(\phi_{i}^{2}+4\phi_{b}^{2}\biggr)\biggr] \\                         
 &\times & \mbox{}   \frac{S}{H^{2}}(S-S_{crit})Im(\sqrt{\xi})=\tilde{I}[\xi_{cl}^{L},\phi_{cl}]
\end{eqnarray*}

The convergence of the integral along this part of the contour is
dependent on the the asymptotic behaviour of the real part of the
Euclidean action on the first quadrant of the  first sheet

\begin{eqnarray*}
Re [I^{I}(\xi,S,\phi _{i},\phi _{b})]&\sim &\biggr[\frac{5}{4\sqrt{2}}+\frac{1}{48\sqrt{2}}\biggl(\frac{m}{H}\biggr)^{2}\biggl(\phi_{i}^{2}+4\phi_{b}^{2}\biggr)\biggr] \\
                                      &\times &\mbox{}\frac{S}{H^{2}}(S-S_{crit}^{I})\sqrt{\mid\xi\mid}
\end{eqnarray*}

When there was no scalar curvature coupling, as in \cite{cris1}, there
was a finite maximum value that $S_{crit}$ could take, no matter what the
values of the parameters $m$ and $\phi_{b}$ were. This guaranteed convergence
of the integral in the first quadrant, from a certain value of $S$, whatever the values of $m$
and $\phi_{b}$. Now the situation is different because $S_{crit}$,
$(\ref{scrit})$, includes a term in  $\eta$ that makes $S_{crit}\rightarrow +\infty$
as $\eta \rightarrow -\infty$.


\section{Semiclassical Approximation}

One of the main requirements on any model is that it yields a
wavefunction that in the late universe predicts a classical
(Lorentzian) spacetime, like the one we experience. Now, a
wavefunction of the universe will predict a classical spacetime where
it is well approximated by the semiclassical approximation associated
with Lorentzian classical solutions. From what we have seen above, the
SD contour passing through the classical Lorentzian solutions satisfies  that condition for large enough $S$.

In our model there are two integration variables $\xi $ and $\phi
_{i}$ and the full wavefunction of the universe  is given by
$(\ref{ppsi} )$. We work under the assumption that $\phi _{i}$ is to
be integrated over real values, and $\xi $ over the complex Riemann
surface of the Euclidean action $I$. We now know that the integral
over $\xi $ can be calculated as a steepest descent $(SD)$ integral
for all the relevant values of $\phi _{i}$, and that the action  peaks
about the classical solutions $\overline{\xi }$.

We can thus  replace $\int _{C_{SD}}d\xi e^{-I}$ by its   semiclassical approximation based on the relevant classical extrema. This will give rise to Laplace type integrals in $\phi _{i}$, when the extrema have real Euclidean action, and to Fourier type integrals in $\phi _{i}$, when the extrema have pure imaginary Euclidean action. These integrals can  then be shown to be dominated by the stationary points of the integrand which coincide with the classical solutions $\overline{\phi}_{i}$, where 

$$ \frac{\partial I[\overline{\xi},\phi _{i}]}{\partial \phi _{i}}\mid _{\phi _{i}=\overline{\phi}_{i}}=0 $$   

This justifies the validity of the semiclassical approximation to the full wavefunction.

So  given the  full wavefunction

\be
\Psi (S,\phi _{b})=\frac{S}{2\pi iH^{2}}\int _{C}d\xi d\phi _{i}e^{-I(\xi,S,\phi _{i},\phi_{b})},
\ee
with an  SD contour associated with real classical
Lorentzian solutions $\{\overline{\xi}_{k}(S,\phi _{b})\}$
 with pure imaginary actions
$I_{k}=i\tilde{I}[\overline{\xi}_{k}(S,\phi _{b});\phi
_{i}]=i\tilde{I}_{k}(S,\phi _{b},\phi _{i})$, the semiclassical
approximation of the wavefunction will be

\begin{eqnarray*}
\Psi _{SC}(S,\phi _{b})&\sim &\int d\phi _{i}\sum_{k}\sqrt{\frac{S^{2}}{2\pi H^{4}\mid
\tilde{I}^{''}[\overline{\xi}_{k}(S,\phi _{b}),\phi _{i}]\mid }}e^{-i[
\tilde{I}(\overline{\xi}_{k}(S,\phi _{b}),\phi _{i})  +\mu _{k}\frac{\pi}{4}]} \\
                       &\sim & \mbox{} \sum_{k}\sqrt{\frac{S^{2}}{2\pi
H^{4}\mid \tilde{I}_{k}^{''}(S,\phi _{b})\mid }}e^{-i[
\tilde{I}_{k}(S,\phi _{b})  +\mu _{k}\frac{\pi}{4}]}
\label{prefac}
\end{eqnarray*}
where $'$ means derivative with respect to $\xi$,
and $\mu _{k}=sgn ( \tilde{I}^{''})$.

When the dominant extrema are real Euclidean solutions
$\{\overline{\xi}_{k}(S,\phi _{b})\}$,  with real Euclidean
 actions,   then after the semiclassical
evaluation of the integral over $\xi$, we are left with  Laplace-type integrals
over $\phi _{i}$ which  are dominated by the contributions
coming from the stationary points of $I[\overline{\xi}_{k}(S,\phi
_{b}),\phi _{i}]$,
which are precisely the classical solutions
$\overline{\phi}_{i}^{k}(S,\phi _{b})$.
So the semiclassical wavefunction will then be

\be
\Psi _{SC}(S,\phi _{b})\sim \sum_{k}\sqrt{\frac{S^{2}}{2\pi
H^{4}\mid I_{k}^{''}(S,\phi _{b})\mid }}e^{-
I_{k}(S,\phi _{b})}
\ee

Since we are mainly interested in knowing if this model predicts
classical Lorentzian spacetime for the late universe, we have computed
the semiclassical wavefunction associated with the  classical Lorentzian
branch of solutions  near  $\xi=1/4$ in figure $2$. We considered  $\eta =0.225, m=1$
and $\phi_{b}=1$, and the result obtained is  shown in figure $8$. The behaviour exhibited
during the late universe, (i.e. large values of $S$) is
typical of that of a wavefunction describing a classical Lorentzian
universe , as desired.

As for the Euclidean solutions, their semiclassical contribution is
exponentially suppressed except for small values of $S$; see figure
$9$. Furthermore, it should be noted that this suppression becomes
increasingly strong as  $\eta$ grows.
The peak in the semiclassical wavefunction is not caused by the
behaviour of the Euclidean action, which is monotonically increasing
as $S$ increases, but by the pre-factor involving the second
derivative of the action. 

It is clear that although there are classical Euclidean solutions for
any value of $S$ the probability associated with a Euclidean universe
with large boundary, 
(large $S$), is very small, and so the late universe wavefunction should
be well approximated by the semiclassical wavefunction associated with
the Lorentzian solutions.

\section{Conclusions}

The addition of the arbitrary scalar coupling $\eta R \phi^{2}$ has
produced results never seen in all of the previous simplicial models
considered in \cite{h2}, \cite{birm}, \cite{cris1} and \cite{cris2}.
First, we note  the existence of real Euclidean classical solutions for any
size of the boundary $three-$space. However, the contribution of these
solutions to the wavefunction of the universe was seen to be
exponentially suppressed except for small boundary $three-$spaces. In the
late universe the wavefunction was found to be dominated by the
contribution of classical Lorentzian solutions.
Second,  the SD contours passing though
these solutions are different from the ones obtained in
\cite{cris1}. The SD path associated to the Lorentzian solution in the
first sheet of Riemann surface of $I[\xi]$, starts off at infinity moving downwards
through  the first quadrant,   ending up precisely at the classical Lorentzian
solution, and does not cross into the second sheet as before. 
Third,
the behaviour of the real part of the action in the first quadrant 
now depends on the value of $\eta$ relative to the value of $S$,
because $S_{crit}\rightarrow +\infty$ as $\eta\rightarrow -\infty$. 
So we see that since the value of $\eta$ is arbitrary  so is the value
$S_{crit}$ from which the SD path converges. 

Nevertheless for any given $\eta$ there is a value of $S$ from which
the SD integral is convergent and is dominated by the contribution of
the Lorentzian classical solutions. Furthermore,  larger values of $S$
lead to a stronger peak in the action around those classical solutions
and this  makes the semiclassical approximations of the SD wavefunction
quite good. The oscillatory behaviour of the semiclassical
wavefunctions indicates that the wavefunction of the universe for this
model predicts classical Lorentzian spacetime for the late universe,
(large $S$).

\end{document}